\documentclass[conference]{IEEEtran}
\IEEEoverridecommandlockouts
\usepackage{cite}
\usepackage{amsmath,amssymb,amsfonts}

\usepackage{algorithm}
\usepackage{algorithmic}
\usepackage{float}
\usepackage{graphicx}
\usepackage{textcomp}
\usepackage{xcolor}
\usepackage{subfigure}
\usepackage{enumitem}
\usepackage{flushend}
\usepackage{hyperref} 
\usepackage{soul}

\def\BibTeX{{\rm B\kern-.05em{\sc i\kern-.025em b}\kern-.08em
    T\kern-.1667em\lower.7ex\hbox{E}\kern-.125emX}}

\begin{document}

\title{
Learning Low-Dimensional Representation for O-RAN Testing via Transformer-ESN}

\author{
    \IEEEauthorblockN{Jiongyu Dai\IEEEauthorrefmark{1}, Raymond Zhao\IEEEauthorrefmark{1}, Farhad Rezazadeh\IEEEauthorrefmark{2}, Lizhong Zheng\IEEEauthorrefmark{3}, Haining Wang\IEEEauthorrefmark{1}, and Lingjia Liu\IEEEauthorrefmark{1}}
    \IEEEauthorblockA{\IEEEauthorrefmark{1}Virginia Tech, Blacksburg, VA, USA \\
    \IEEEauthorblockA{\IEEEauthorrefmark{2}Universitat Politècnica de Catalunya (UPC), Barcelona, Spain}
    \IEEEauthorblockA{\IEEEauthorrefmark{3}Massachusetts Institute of Technology, Boston, Massachusetts, USA}
    Email:\{jiongudai, zraymond, hnw, ljliu\}@vt.edu, farhad.rezazadeh@upc.edu, lizhong@mit.edu}
    \thanks{This paper has been accepted for publication at IEEE International Conference on Mobile Ad-hoc and Smart Systems (MASS) 2025, doi: \href{https://doi.org/10.1109/MASS66014.2025.00030}{10.1109/MASS66014.2025.00030}. © 2025 IEEE.  Personal use of this material is permitted.  Permission from IEEE must be obtained for all other uses, in any current or future media, including reprinting/republishing this material for advertising or promotional purposes, creating new collective works, for resale or redistribution to servers or lists, or reuse of any copyrighted component of this work in other works.}
}

\maketitle

\begin{abstract}
Open Radio Access Network (O-RAN) architectures enhance flexibility for 6G and NextG networks.
However, it also brings significant challenges in O-RAN testing with evaluating abundant, high-dimensional key performance indicators (KPIs). 
In this paper, we introduce a novel two-stage framework to learn temporally-aware low-dimensional representations of O-RAN testing KPIs. 
To be specific, stage one employs an information-theoretic H-score to train a hybrid self-attentive transformer and echo state network (ESN) reservoir, called Transformer-ESN, capturing temporal dynamics and producing task-aligned $8$-dimensional embeddings. 
Stage two evaluates these embeddings by training a lightweight multilayer perceptron (MLP) predictor exclusively on them for key target KPIs such as reference signal received quality (RSRQ) and spectral efficiency. 
Using real-world O-RAN testbed data (video streaming with interference), our approach demonstrates a significant advantage specifically when training samples are very limited. 
In this scenario, the low-dimensional representations learned from the Transformer-ESN yield mean square error (MSE) reductions of up to 41.9\% for RSRQ and 29.9\% for spectral efficiency compared to predictions from the original high-dimensional data.
The framework exhibits high efficiency for O-RAN testing, significantly reducing testing complexities for O-RAN systems.

\end{abstract}

\begin{IEEEkeywords}
O-RAN, Low-Dimensional Representation, H-score, Transformer, Echo State Network (ESN)
\end{IEEEkeywords}

\section{Introduction}
\label{sec:intro}

The evolution towards 6G networks increasingly relies on architectures like the open radio access network (O-RAN), which promises enhanced flexibility, vendor interoperability, and innovation through its disaggregated and standardized interfaces \cite{singh2020evolution, ShadabNTNO-RAN}. Realizing this vision, however, introduces significant operational complexity. The O-RAN Alliance's testing specifications mandate the monitoring of numerous key performance indicators (KPIs) across various applications, network layers, and functional components \cite{oran-e2e}. This extensive monitoring is crucial for tracking network performance and ensuring that diverse end-user services, particularly demanding applications like real-time video streaming, meet their required quality of service (QoS) levels.


This mandated abundance of high-dimensional KPI data presents a critical challenge in terms of scalability and efficiency. The sheer volume of metrics, collected frequently from potentially thousands of network elements, creates a combinatorial explosion for testing and validation. It significantly increases data storage and transmission overhead, especially towards management layers like the service management and orchestration (SMO) framework. Furthermore, extracting timely, actionable insights from this "data deluge" for network optimization and troubleshooting becomes exceedingly difficult. Many of these monitored KPIs are often highly correlated or redundant, containing overlapping information about the underlying network state. As O-RAN deployments scale towards 6G, efficiently managing this vast amount of potentially redundant data is not merely an operational convenience but a critical requirement for sustainable and cost-effective network rollout and operation.

To tackle this challenge, we propose shifting from monitoring exhaustive raw KPI sets to learning compact, low-dimensional embeddings that capture the essential state of the network and its ability to deliver services effectively. The core idea is that the underlying system dynamics and performance variations, even across many KPIs, are often driven by a smaller set of latent factors. Many KPIs may be redundant, offering little additional predictive value beyond a core subset. By learning an efficient mapping from the high-dimensional observed KPI space to a low-dimensional latent space, we aim to create representations that retain the most critical predictive information while discarding redundancy, thereby facilitating more scalable monitoring, analysis, and control.

In this work, we develop and evaluate a novel two-stage framework specifically designed to learn such compact, temporally-aware representations from O-RAN KPI sequences. The first stage focuses on feature extraction using an information-theoretic objective, the H-score \cite{xxjmlr}, to guide the learning process towards task-relevant features. We employ a hybrid self-attentive transformer and echo state network (ESN) reservoir architecture to effectively capture both long-range dependencies and fine-grained temporal dynamics inherent in KPI streams. This stage produces task-aligned embeddings that significantly reduce the high-dimensional input sequence into low-dimensional representations. The second stage evaluates the utility of these learned embeddings by training a lightweight multilayer perceptron (MLP) predictor exclusively on them to predict key target KPIs – specifically, reference signal received quality (RSRQ) and spectral efficiency for a video streaming use case. The prediction accuracy serves as a quantitative measure of how well our low-dimensional mapping captures the relevant system state, demonstrating its potential for efficient service-level assessment. We validate our approach using data collected from a real-world O-RAN testbed and show that our framework yields strong generalization and efficiency gains. Specifically, our contributions can be summarized as the following: 


\begin{itemize}
    \item Development of an O-RAN-compliant testbed incorporating SDR-based UEs/BS, a custom interferer, and video streaming services for collecting realistic KPI datasets.

    \item Application of an information-theoretic learning approach to train deep learning models for extracting compact, task-aligned representations from high-dimensional O-RAN KPI sequences

    \item Design a novel hybrid Transformer-ESN architecture tailored for capturing relevant temporal patterns and dependencies in KPI time series data. 

    \item Demonstration of efficient and accurate KPI prediction using the learned low-dimensional mappings, showing significant advantages over baseline models in data-limited scenarios.
\end{itemize}

The paper is organized as follows. Section~\ref{sec:relatedwork} discusses related work. Section~\ref{sec:systemmodel} presents the system model, followed by our methodology in Section~\ref{sec:method}. Section~\ref{sec:experiment} details the experimental setup and data collection. Section~\ref{sec:results} presents the performance evaluation. Section~\ref{sec:concl} concludes this paper.
\section{Related Work}
\label{sec:relatedwork}

\textbf{O-RAN System Monitoring and Challenges:} The disaggregated nature of O-RAN creates significant monitoring challenges due to the high volume of high-dimensional, often redundant data, impacting overall system scalability. Machine learning techniques have been applied to enhance specific O-RAN analyses, such as traffic classification using frameworks like TRACTOR~\cite{gro2024} or video quality prediction via multimodal integration as in DeepQoE~\cite{zhang2020deepqoe}. In \cite{ShadabNTNO-RAN}, an AI-based prediction network is discussed to cope with the delayed observation of network states. Concurrently, managing operational aspects like scaling the near-RT RIC under heavy xApp load is addressed by solutions such as ORAN-HAutoscaling~\cite{info16040259}, which tackles performance bottlenecks arising from system complexity. However, while these studies provide valuable solutions for specific applications or operational aspects, a defined, general approach for efficiently monitoring and extracting insights from the multitude of KPIs across increasingly complex and large-scale O-RAN deployments remains an open challenge.


\textbf{Dimensionality Reduction for Network Data:} Given the high dimensionality and potential redundancy in O-RAN KPIs, dimensionality reduction is crucial for efficiency. Common approaches include feature selection or pruning, aiming to identify and retain only the most informative KPIs. Techniques range from explainability-driven methods \cite{tassie2024shap} to classic feature selection algorithms like RReliefF or recursive feature elimination \cite{stojcic2023lte}. While effective in trimming KPI sets for specific modeling tasks, these methods often operate as static preprocessors and may not adapt well to dynamic network conditions or integrate seamlessly into end-to-end learning frameworks. An alternative is to learn compact, task‑aware representations using neural networks. For instance, \cite{polese2022colo} employs an Autoencoder to map raw KPIs into a low‑dimensional feature vector, which then feeds a deep‑reinforcement‑learning (DRL) agent for network‑slicing decisions. 

\textbf{Sequence Modeling Architectures:} Modeling KPI sequences requires architectures adept at capturing temporal dependencies. ESNs, a type of Reservoir Computing, offer an efficient approach. 
Their suitability for wireless tasks has been demonstrated in applications like MIMO symbol detection \cite{zzhou2020} and intelligent radio resource management \cite{chang2020deep,jdy2025}. Furthermore, ESNs have been specifically explored for multivariate time series prediction, with architectural variations like parallel and multiscale reservoirs to improve feature extraction capabilities for complex system dynamics\cite{li2024multi}. Further extending RC approaches, \cite{xjr-attention} integrate attention mechanisms and decision feedback with reservoir computing for improved online MIMO-OFDM symbol detection. 

The Transformer architecture, introduced initially for natural language processing \cite{vaswani2017attention}, has also become highly influential in sequence modeling tasks. 
Additionally, its application has extended successfully to time series analysis, demonstrating strong performance in tasks such as long sequence forecasting \cite{zhou2021informer} and multivariate time series representation learning \cite{zerveas2021transformer}. 
Our work, on the other hand, seeks to combine the strengths of both ESNs and Transformers in a hybrid architecture tailored for O-RAN KPI analysis.

\section{System Model}
\label{sec:systemmodel}

\begin{figure*}[h]
    \centering
    \includegraphics[width=0.8\linewidth]{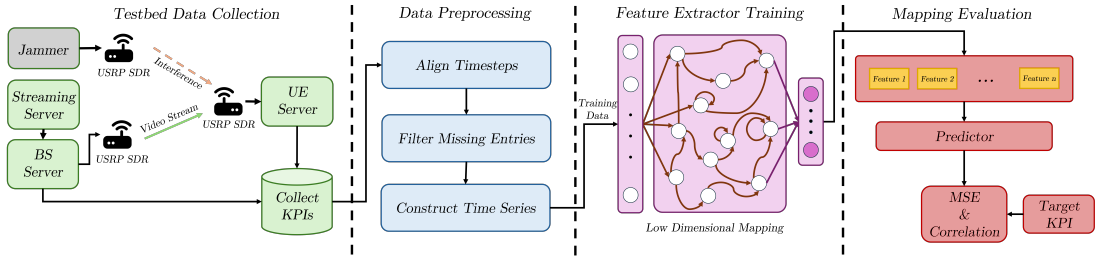}
    \caption{System flow diagram}
    \label{fig:sys_block}
\end{figure*}

We consider a general setting in which a high-dimensional input $\mathbf{X} \in \mathbb{R}^{N_{\text{seq}}\times K}$ defines a sequence of observed O-RAN KPIs, and the target variable $\mathbf{Y} \in \mathbb{R}^{K}$ represents a set of performance metrics or quality-of-service (QoS) indicator. The objective is to learn a low-dimensional representation $f(\mathbf{X}) \in \mathbb{R}^n$, with $n \ll N_{\text{seq}} \times K$, that captures the information in $\mathbf{X}$ most relevant to characterizing or predicting $\mathbf{Y}$.

Since the true joint distribution $P_{\mathbf{X},\mathbf{Y}}$ is unknown, we aim to discover a function $f : \mathbb{R}^{K} \to \mathbb{R}^n$ that approximates the statistical dependence between $\mathbf{X}$ and $\mathbf{Y}$. To support this alignment, we also introduce a matching transformation $g : \mathbb{R}^{K} \to \mathbb{R}^n$ that maps the target variable into the same latent space, ensuring that comparisons between $f(\mathbf{X})$ and $g(\mathbf{Y})$ are dimensionally consistent during training. Ideally, the learned representation $f(\mathbf{X})$ should act as a low-dimensional sufficient statistic of $\mathbf{X}$ with respect to $\mathbf{Y}$.

We formalize the learning objective as the following variational optimization problem:
\begin{equation}
    f^* = \text{arg}\min_{f \in \mathcal{F}_n} \mathcal{L}(f),
\end{equation}
where $\mathcal{F}_n$ denotes a class of $n$-dimensional feature maps and $\mathcal{L}(f)$ is a substitute loss function measuring statistical misalignment between the extracted features $f(\mathbf{X})$ and target structure in $\mathbf{Y}$.
To avoid relying on mutual information—which is difficult to estimate accurately in high-dimensional settings—we instead use empirical proxies that encourage statistical dependence between $f(\mathbf{X})$ and $g(\mathbf{Y})$. A general form of the objective can be shown as:
\begin{equation}
    \mathcal{L}(f) = \mathbb{E}_{\mathbf{X},\mathbf{Y}} \left[ \ell\big(f(\mathbf{X}), g(\mathbf{Y})\big) \right],
\end{equation}
where $\ell(\cdot, \cdot)$ penalizes divergence between the learned features and relevant structure in the target.



\section{Methodology}
\label{sec:method}


The overarching goal of our system is to extract compact, task-relevant representations from high-dimensional O-RAN KPI sequences to support efficient prediction and evaluation. As such, we design a modular two-stage pipeline that systematically applies dimensionality reduction, predictive modeling, and evaluation on preprocessed testbed data. Figure~\ref{fig:sys_block} provides a visual overview of this entire workflow, starting from data collection in our O-RAN testbed, followed by data preprocessing, training of the feature extractor, and finally evaluating the learned low-dimensional mapping with a prediction task.

The full pipeline operates as follows:
\begin{enumerate}
    \item \textbf{Feature Extraction:} We first train a neural feature extractor to transform sequences of high-dimensional KPI data into a low-dimensional mapping. This extractor is built using a \emph{Transformer--ESN hybrid network}, designed to capture both long-range dependencies and dynamic temporal patterns. Crucially, the training objective is not conventional supervised prediction, but rather the \emph{H-score} metric, which quantifies statistical alignment between input sequences and target KPI vectors. This metric guides the model to learn compact latent embeddings that are maximally predictive.

    \item \textbf{Evaluation via Prediction:} The effectiveness of the learned representation is then evaluated on a held-out test set. We first freeze the parameters of the trained feature extractor and use it to encode the input dataset into the low-dimensional embeddings. A simple MLP predictor is then trained exclusively on these embeddings to estimate target KPIs. The prediction accuracy of this MLP, operating solely in the reduced space, reflects how much relevant structure was preserved by the mapping. Finally, we benchmark this performance against a baseline model trained directly on the original input data, using standard regression metrics (results in Section~\ref{sec:results}).
 
\end{enumerate}

The following subsections describe the key components of our system, focusing on the information-theoretic objective and neural architecture used to extract low-dimensional representations from high-dimensional KPI sequences.

\subsection{Feature Learning and H-score}

Although many performance metrics are observed during O-RAN testing, they are often driven by a smaller set of latent variables that govern overall system behavior. Our goal is to uncover these low-dimensional, task-relevant structures by learning a low-dimensional representation of the input that preserves its statistical relationship with the target. Ideally, this representation, generated by mapping functions, contains all the information about the target within the input, indicating that the mapping functions should be sufficient statistics. Thus, learning the low-dimensional representation can be formulated as a problem of identifying sufficient statistics. 
To address this, we follow the information-theoretic framework in \cite{xxjmlr}.

Let $\mathcal{F}_X = \{f : \mathcal{X} \rightarrow \mathbb{R}\}$ denote the space of real-valued feature functions over the input domain. We wish to find a set of functions $[f_1(X), \ldots, f_n(X)]$ that form a sufficient statistic: a minimal collection that retains all the information in $X$ relevant for predicting $Y$. Within a neural network, these functions correspond to hidden-layer activations that act as informative representations.
To formalize this, we consider the pointwise mutual information (PMI), to quantify dependence between $X$ and $Y$:
\begin{equation}
    \text{PMI}(x,y)= \log \frac{P_{XY}(x,y)}{P_X(x)P_Y(y)}, \quad x\in\mathcal{X}, \ y\in\mathcal{Y},
    \label{eq:pmi}
\end{equation}
where $P_X$ and $P_Y$ denote the marginal distributions of $X$ and $Y$ respectively, and $P_{XY}$ represents their joint distribution. If we could write $\text{PMI}(x, y)$ exactly as a sum of separable feature products $f_i(x)g_i(y)$, then the pair $\{f_i, g_i\}$ would form a sufficient representation of the dependency structure. This motivates the decomposition:
\begin{equation}
    \text{PMI}(x, y) = \sum_{i=1}^n (f_i \otimes g_i)(x, y),
    \label{eq:decompo}
\end{equation}
where $(f \otimes g)(x, y) \triangleq f(x) g(y)$. It can be shown that the collection of $f_i$ is the sufficient statistic if this decomposition is achieved, meaning that this set of functions is the optimal mapping function that could be learned to generate the low-dimensional representation.
For high-dimensional data, however, an exact decomposition when $n$ is limited is usually infeasible. Instead, we aim to find the best approximation by minimizing the error:
\begin{equation}
\label{eq:min-pmi}
    \min_{\{f_i, g_i\}_{i=1}^n} \| \text{PMI} - \sum_{i=1}^n f_i \otimes g_i \|^2.
\end{equation}

To make this optimization tractable, we adopt the H-score~\cite{xxjmlr} as a data-driven surrogate objective. It replaces the unknown PMI with empirical second-order statistics:
\begin{align}
    \mathcal{H}(f, g) &\triangleq \sum_{i=1}^{n} \mathrm{cov}(f_i, g_i) 
    - \frac{1}{2} \sum_{i,j=1}^{n} \mathbb{E}[f_i f_j] \cdot \mathbb{E}[g_i g_j] \label{eq:hscore_def} \\
    &= \mathrm{tr}(\mathrm{cov}(\bar{f}, \bar{g})) 
    - \frac{1}{2} \, \mathrm{tr}\left( \mathbb{E}[\bar{f} \bar{f}^T] \cdot \mathbb{E}[\bar{g} \bar{g}^T] \right), \label{eq:hscore_trace}
\end{align}
where $\bar{f} = [f_1(X), \ldots, f_n(X)]^T$ and $\bar{g} = [g_1(Y), \ldots, g_n(Y)]^T$. This objective satisfies $\mathcal{H}(f, g) \cong \|\text{PMI}\|^2 - \| \text{PMI} - \sum_{i=1}^n f_i \otimes g_i \|^2$,
so maximizing the H-score is approximately equivalent to minimizing the PMI approximation error.

\begin{figure}[t!]
    \centering
    \includegraphics[width=.7\linewidth]{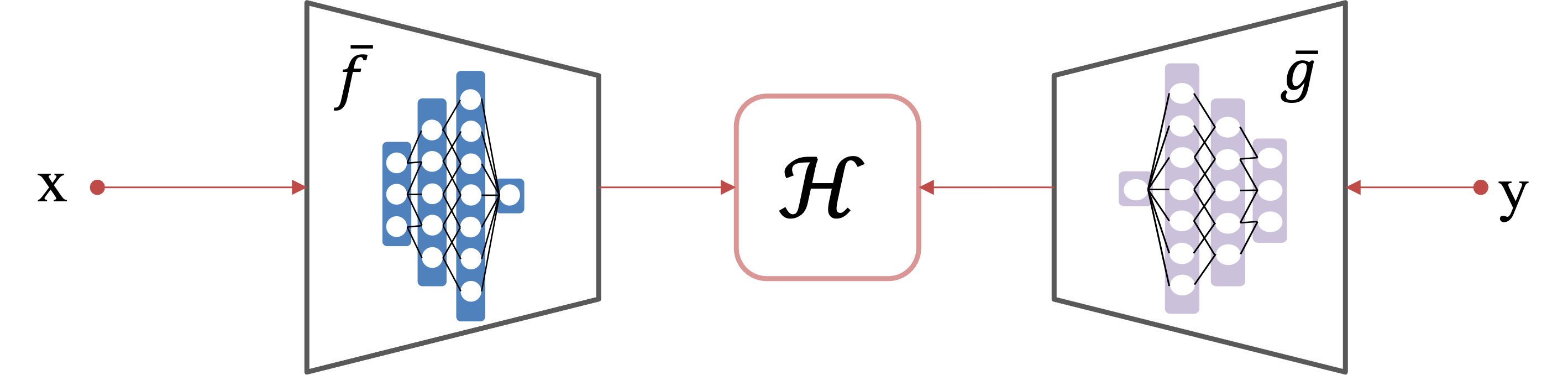}
    \caption{The illustration of H-score network.}
    \label{fig:h-score}
\end{figure}

In practice, $\bar{f}(X)$ and $\bar{g}(Y)$ are parameterized by two neural networks: a Transformer–ESN hybrid for the input and an MLP for the target, represented as blocks in Figure~\ref{fig:h-score}. During training, batches of paired data $(x, y)$ are passed through the respective networks, the H-score is computed over the batch, and its negative is used as a loss function for backpropagation.

By restricting the output dimension of $\bar{f}$ to $n \ll K$, the learned representation serves as a compact surrogate for the full KPI input, optimized to retain structure relevant to $Y$. In our case, $X$ corresponds to a historical KPI sequence and $Y$ to a future KPI vector, allowing the representation to encode dynamics that are predictive of system evolution.

\subsection{Hybrid Transformer--ESN Network Architecture}

The core of our approach is a hybrid feature extraction model that combines a Transformer encoder with an ESN (T-ESN), enabling rich temporal representation from KPI sequences with minimal tuning.

\textbf{Transformer Encoder.}  
Given an input sequence $\{\mathbf{x}_t\}_{t=1}^{N_{\text{seq}}}$, where $\mathbf{x}_t \in \mathbb{R}^{K}$ represents KPIs at time $t$, each vector is first projected into a higher-dimensional space via a learned linear layer with added positional encoding:
\begin{equation}
    \mathbf{z}_t = W_{\text{proj}} \mathbf{x}_t + \mathbf{b}_{\text{proj}} + \mathbf{p}_t.
\end{equation}
Stacking these yields $\mathbf{Z} \in \mathbb{R}^{N_{\text{seq}} \times d_{\text{model}}}$, which is passed through $L$ Transformer layers, each consisting of multi-head self-attention and feedforward blocks, following the well-defined architecture from~\cite{vaswani2017attention}.

\textbf{Reservoir Encoding.}  
The final Transformer output $\mathbf{U} = [\mathbf{u}_1, \ldots, \mathbf{u}_{N_{\text{seq}}}]$ is passed through a projection layer and used to drive an ESN, which models temporal dynamics via its recurrent reservoir. The state update is given by:
\begin{equation}
    \mathbf{s}_t = \sigma(W_{\text{res}} \mathbf{s}_{t-1} + W_{\text{in}} \mathbf{u}_t),
\end{equation}
where $\sigma(\cdot)$ is a nonlinear activation (e.g., $\tanh$). The final ESN state $\mathbf{s}_{N_{\text{seq}}}$ is concatenated with the flattened original input to form the trainable readout vector, $W_{\text{out}}$:
\begin{equation}
    \mathbf{y} = W_{\text{out}} [\mathbf{s}_{N_{\text{seq}}}; \mathbf{x}_1; \ldots; \mathbf{x}_{N_{\text{seq}}}],
\end{equation}
which serves as the low-dimensional output $f(\mathbf{X}) \in \mathbb{R}^n$ used for H-score training. Based on empirical evaluation balancing dimensionality reduction and predictive performance, we chose to set the target dimension \(n=8\).

\begin{figure}[t!]
    \centering
    \includegraphics[width=0.9\linewidth]{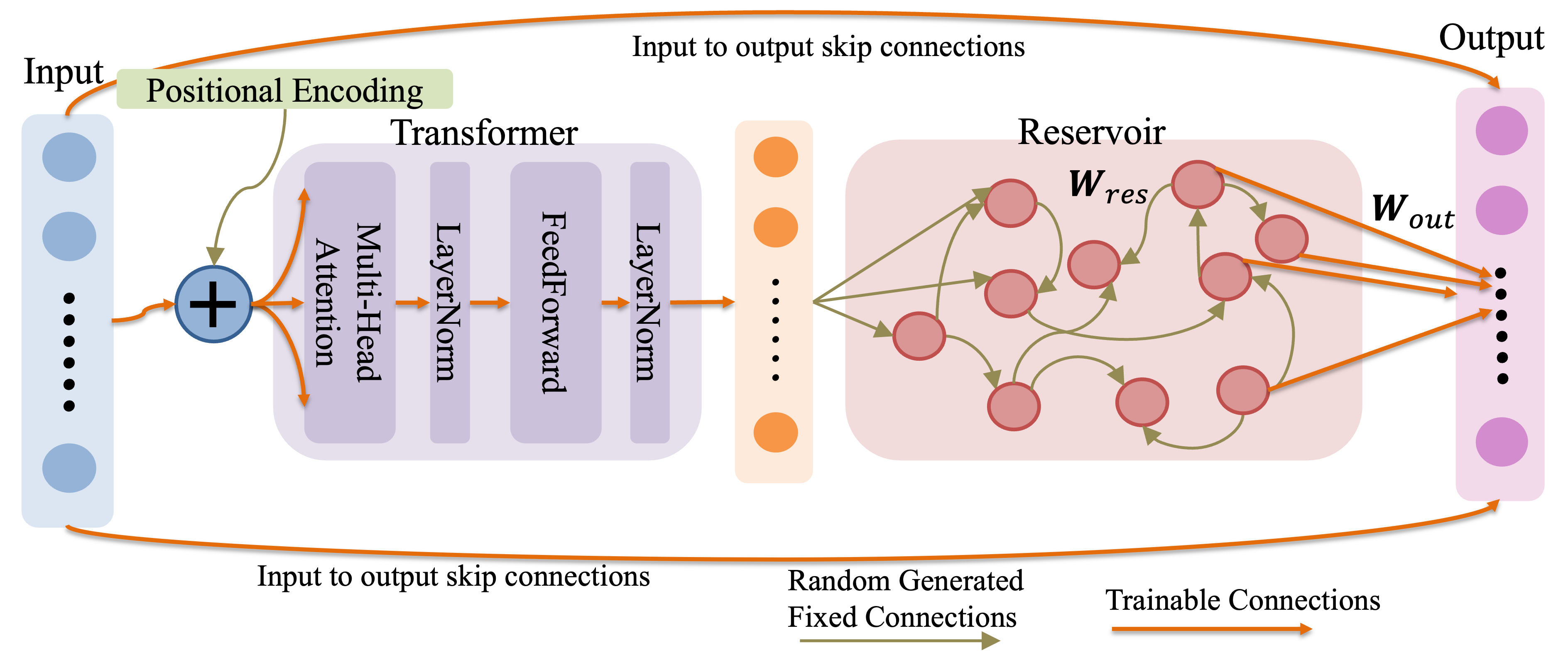}
    \caption{Hybrid Transformer--ESN feature extraction pipeline.}
    \label{fig:tesn}
\end{figure}

The overall network architecture of T–ESN is visualized in Figure~\ref{fig:tesn}, illustrating the Transformer layers feeding into the ESN reservoir, and the combined readout supporting supervised learning and statistical feature extraction.

\section{Experiments}
\label{sec:experiment}

This section describes our O-RAN testbed, the data preprocessing pipeline, and the final dataset used for evaluation.

\subsection{Testbed Setup}

\begin{figure}[h!t]
    \centering
    \subfigure[Cloud-based testbed on CCI xG platform]{
     \begin{minipage}[t]{.8\linewidth}
      \centering
      \includegraphics[width=\linewidth]{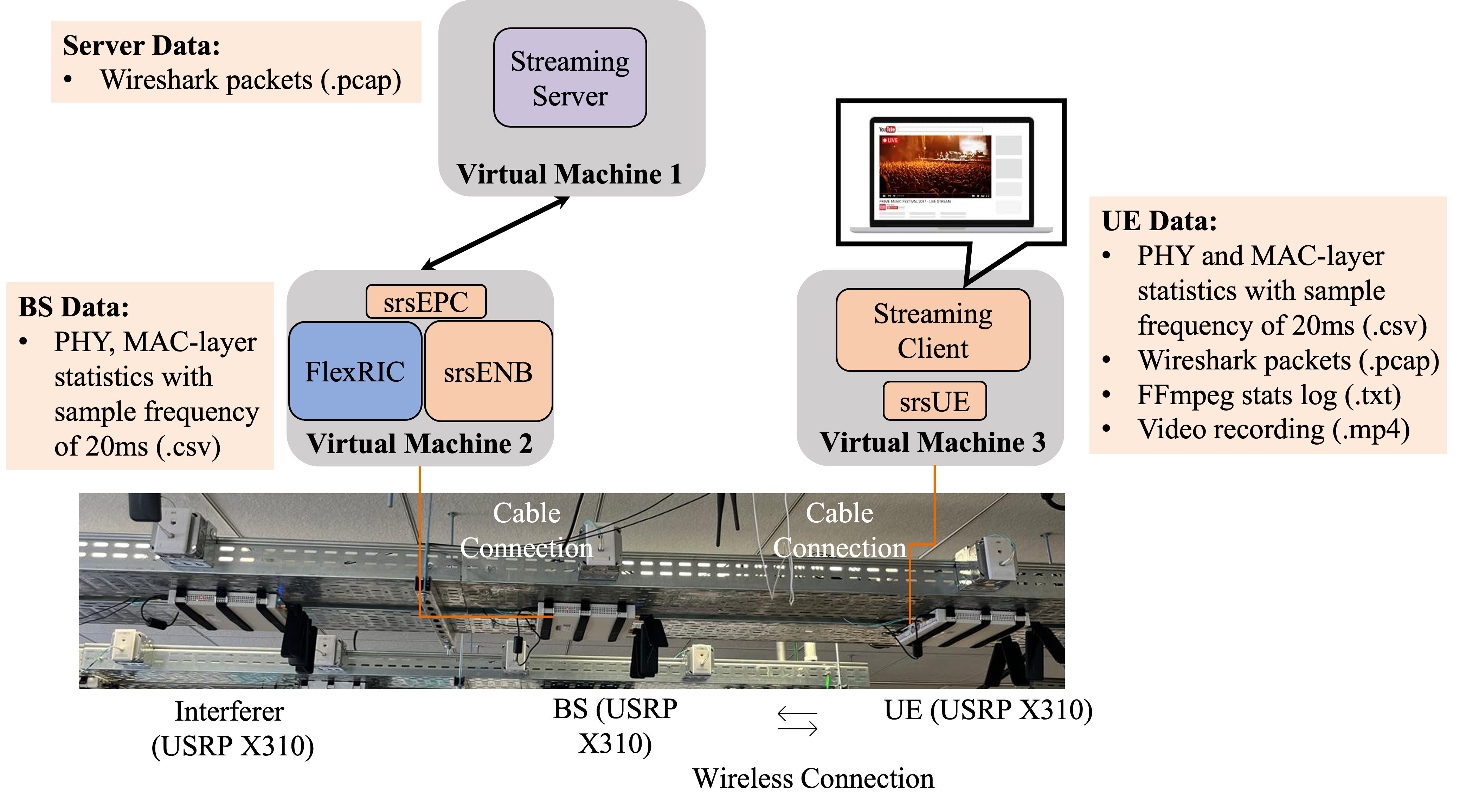} 
      \label{fig:cci_testbed}
     \end{minipage}
    }
    \subfigure[Local testbed with B210 USRPs]{
     \begin{minipage}[t]{.8\linewidth}
      \centering
      \includegraphics[width=\linewidth]{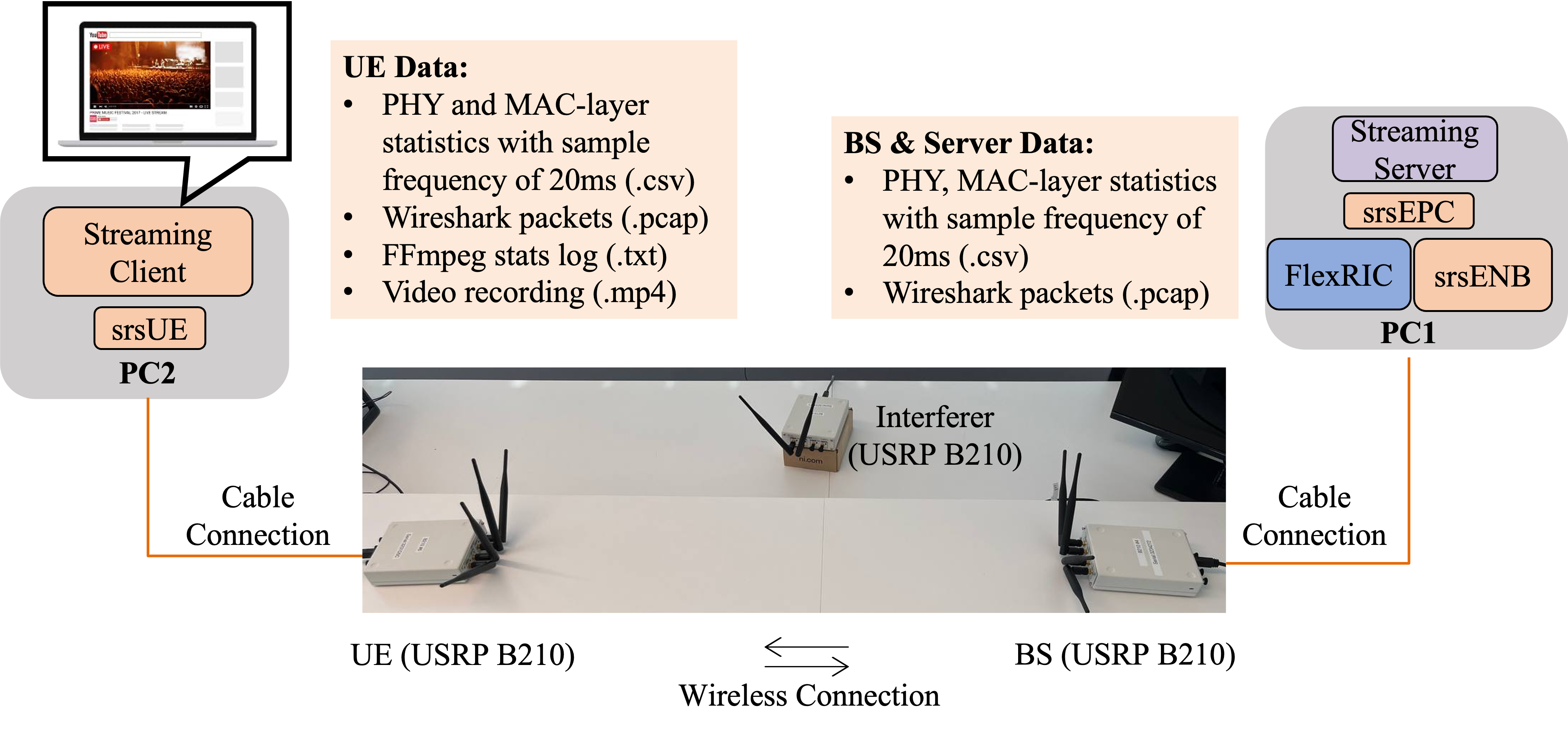} 
      \label{fig:lab_testbed}
     \end{minipage}
    }
\caption{O-RAN testbed deployments used for KPI data collection.}
\label{testbed}
\end{figure}

Our testbed includes a near-real-time RAN intelligent controller (RIC), base station (BS), user equipment (UE), video streaming server, and interferer. The RIC uses FlexRIC from the MOSAIC5G project, and the video pipeline is implemented with MediaMTX and FFmpeg.
The BS and UE are deployed using srsRAN on USRP SDRs. The BS transmits at 2680 MHz downlink across 25 resource blocks in 2$\times$2 MIMO FDD. A custom interferer, implemented in C++ with UHD, introduces interference via random OFDM bursts and varying gain/sleep durations. Each 120-second run logs PHY/MAC-layer KPIs, packet captures, and video stats every 20 ms.
We deploy two variants of this setup (Figure~\ref{testbed}): one on the CCI xG testbed with USRP X310s and OpenStack virtualization, and one on a local testbed with B210s and standalone Ubuntu hosts.




\subsection{Data Preprocessing} \label{data_processing}

Following raw data collection, we perform a series of preprocessing steps to produce aligned, filtered, and temporally structured KPI sequences suitable for model training. These steps include moving average smoothing, missing value handling, and construction of fixed-length input–target pairs.

\subsubsection{Moving Average Aggregation}

To align raw measurements collected from different components,
we apply a fixed-length time window to compute a moving average. For each KPI, values falling within a given window are averaged and assigned the window's start time \( t_{\text{start}} \) as the new timestamp. The window then advances by a fixed step \( t_{\text{step}} \), and the process repeats over the duration of the experiment.
This yields a uniformly spaced sequence of averaged KPI values for each experiment run. The resulting sequences are compiled into the dataset \( D_{\text{kpi}} \), with each row corresponding to one time step. 

\subsubsection{Padding and Row Filtering}


To prepare the dataset, we handle missing and anomalous values in \( D_{\text{kpi}} \). Since packet transmission isn't guaranteed in every interval (e.g., awaiting acknowledgments), a missing 'UE Packet Delay' often signifies the absence of a relevant packet event rather than a data error, especially if other KPIs are present. Therefore, rows missing \textit{only} this value are retained, with the field filled in as \(-1\) to mark this specific absence. However, if any other KPI is missing, suggesting a broader data issue for that timestamp, the entire row is dropped to ensure data quality.

We also remove outliers by applying interquartile range (IQR) filtering with thresholds defined at the 10th and 90th percentiles. Values beyond the range \([Q_1 - 1.5 \times IQR,\, Q_3 + 1.5 \times IQR]\) are removed.



\subsubsection{Constructing Sequential Data}

To prepare the input–target pairs for training, we extract continuous KPI sequences of length \( N_{\text{seq}} \) from \( D_{\text{kpi}} \). We set \( N_{\text{seq}} = 28\), chosen to capture relevant short-to-medium term temporal dependencies, across $K=13$ KPI features. For each candidate row \( i \), we verify that all preceding \( N_{\text{seq}} - 1 \) rows have consecutive timestamps spaced by \( t_{\text{step}} \). If valid, the $28\times13$ sequence ending at row \( i \) is formed into an input sample $X$. The corresponding KPI vector from row \( i + 1 \) is taken as the sample $Y$. Repeating this process for all valid starting rows yields the final datasets $D_x=\{X_j\}_{j=1}^{M}$ and $D_y=\{Y_j\}_{j=1}^{M}$

\subsection{Dataset Overview}

The selection of KPIs in our dataset is guided by the O-RAN Alliance's End-to-End Test Specification~\cite{oran-e2e}, which outlines metrics relevant to video streaming applications. This ensures our measurements are aligned with standardized testing methodologies for evaluating O-RAN system performance.
After running the testbed experiments and collecting logs from multiple sources (e.g., pcap traces, CSV files, FFmpeg logs), we applied the preprocessing pipeline described in Section~\ref{data_processing} to construct the final KPI dataset. Table~\ref{tab:kpis} shows the empirical statistics of the KPI data collected with the explanation of what each KPI measures.
The resulting dataset contains a total of 59{,}441 samples, with 23{,}776 used for training and 35{,}665 held out for testing. The dataset is publicly available at \cite{orandataset}.

\begin{table}[]
\centering
\caption{Descriptive statistics for KPIs collected in the tested experiments.}
\label{tab:kpis}
\resizebox{.9\linewidth}{!}{%
\begin{tabular}{c|c|c|c|c}
\hline
\textbf{KPI} &
  \textbf{Unit} &
  \textbf{Description} &
  \textbf{Empirical Min–Max} &
  \textbf{Empirical Mean $\pm$ Std} \\ \hline
\textbf{Spectral Efficiency} &
  bps/Hz &
  Downlink throughput (bps) divided by the channel bandwidth\cite{oran-e2e} &
  0.00 - 3.74 &
  0.58 $\pm$ 0.38 \\
\textbf{RSRP} &
  dBm &
  Average per‑resource‑element reference‑signal power seen by UE &
  -102 - -75 &
  -87.58 $\pm$ 3.70 \\
\textbf{SINR} &
  dB &
  Signal‑to‑interference‑plus‑noise ratio measured by UE &
  9.43 - 24.33 &
  18.31 $\pm$ 1.92 \\
\textbf{MIMO Rank} &
  — &
  Number of spatial layers scheduled &
  1 - 2 &
  1.36 $\pm$ 0.38 \\
\textbf{MCS} &
  index &
  Transport‑block modulation‑and‑coding scheme &
  0 - 27 &
  9.04 $\pm$ 4.93 \\
\textbf{RB Number} &
  RBs &
  Physical resource blocks allocated for downlink transmission &
  2 - 25 &
  22.31 $\pm$ 4.34 \\
\textbf{CQI} &
  index &
  Wideband Channel Quality Indicator fed back by UE &
  0 - 13 &
  8.51 $\pm$ 0.92 \\
\textbf{RSRQ} &
  dB &
  Quality metric combining RSRP and total RSSI &
  -14.00 - -6.40 &
  -10.55 $\pm$ 2.47 \\
\textbf{PMI} &
  index &
  Preferred precoder selected by UE &
  0 - 3 &
  0.93 $\pm$ 0.84 \\
\textbf{UE RSSI} &
  dBm &
  Total received signal power level&
  -70 - -60 &
  -65.36 $\pm$ 2.62 \\
\textbf{UE Buffer Status} &
  bytes &
  The amount of data in the UE buffer to be sent out &
  0 - 2944 &
  25.61 $\pm$ 85.34 \\
\textbf{Packet Delay} &
  ms &
  End‑to‑end time between packet at server and client &
  0- 3048.06 &
  62.70 $\pm$ 208.87 \\
\textbf{BLER} &
  \% &
  Block‑error‑rate &
  0 - 78.00 &
  2.64 $\pm$ 6.76 \\ \hline
\end{tabular}%
}
\end{table}

\section{Results Presentation}
\label{sec:results}

In this section, we evaluate the effectiveness of the learned low-dimensional representation \(f(X)\) by assessing its utility in a time-series prediction task. The core idea is to measure how well the low-dimensional representation retains information critical for predicting future KPI values.

\subsection{Evaluation Setup} 

We compare the predictive performance of models trained under different input conditions:
\begin{enumerate}
    \item \textbf{Full KPI + MLP:} A conventional MLP-based predictor trained directly on the \textit{original high-dimensional} KPI sequences. The network has three fully connected layers with 16, 32, and 1 neurons and the ReLU activation function. The same MLP architecture is employed as the predictor across all other tested methods.
    \item \textbf{Autoencoder + MLP:} The encoder—built on the same hybrid Transformer + ESN structure—maps the sequence into an 8‑dimensional feature vector. During training, this vector is fed to a lightweight decoder—a three‑layer fully connected network (16, 32, and 2 neurons, ReLU activations)—which reconstructs the RSRQ and spectral‑efficiency values of the latest KPI vector in the sequence. The decoder is used only for representation learning and is discarded after training. The predictor is the same MLP network mentioned before.   
    \item \textbf{H-score ESN + MLP:} The 8-dimensional features are generated by an ESN trained using the H-score objective. The predictor used in the evaluation is an MLP network trained \textit{only} on these low-dimensional features.
    \item \textbf{H-score T-ESN + MLP:} The 8-dimensional features are generated by the hybrid Transformer-ESN trained using the H-score objective, followed by an MLP predictor trained \textit{only} on these low-dimensional features.
\end{enumerate}

For the H-score methods (items 3 and 4), the feature extractor, \(f(X)\), is trained once using the H-score objective with a three-layer fully-connected network (16, 32, and 8 neurons) as \(g(Y)\). Its parameters are then frozen, and this mapping \(f(X)\) is used to generate embeddings. Separate simple MLP predictors are then trained on these embeddings to predict the two target KPIs: RSRQ and spectral efficiency. This setup ensures the learned representation \(f(X)\) is general and not overly specialized to a single prediction target. To quantify how well these embeddings capture relevant information, the accuracy of the MLP predictors is measured on a held-out test set using standard regression metrics: MSE, which penalizes errors quadratically, and Pearson correlation, which measures the linear correlation between predicted and actual values.

\subsection{Embedding Dimension Selection}

The key objective of this work is to reduce the high dimensionality of O-RAN KPI input sequences while retaining adequate predictive power. Each input sample $X$ consists of a $(28\times 13)$ matrix, representing $N_{\text{seq}}=28$ time steps across $K=13$ KPIs. We aim to learn a mapping $f(X)$ onto a significantly lower-dimensional latent space $\mathbb{R}^n$. Selecting the optimal $n$ involves balancing the degree of reduction in dimension with the potential loss of relevant information necessary for accurate prediction. 

\begin{figure}[h!t]
    \centering
    \subfigure[RSRQ]{
     \begin{minipage}[t]{0.45\linewidth}
      \centering
      \includegraphics[width=\linewidth]{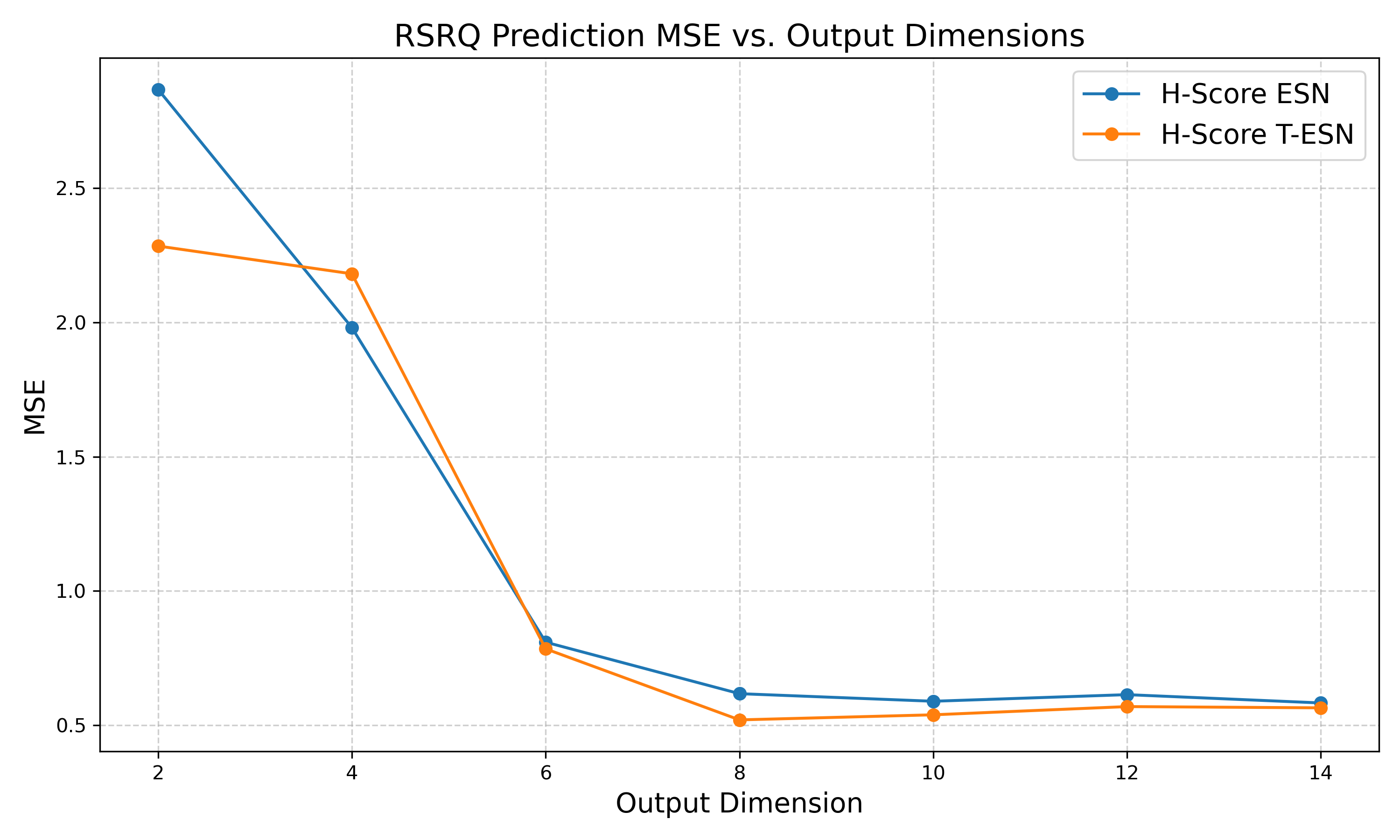} 
      \label{fig:mse_vs_dim}
     \end{minipage}
    }
    \subfigure[Spectral efficiency]{
     \begin{minipage}[t]{0.45\linewidth}
      \centering
      \includegraphics[width=\linewidth]{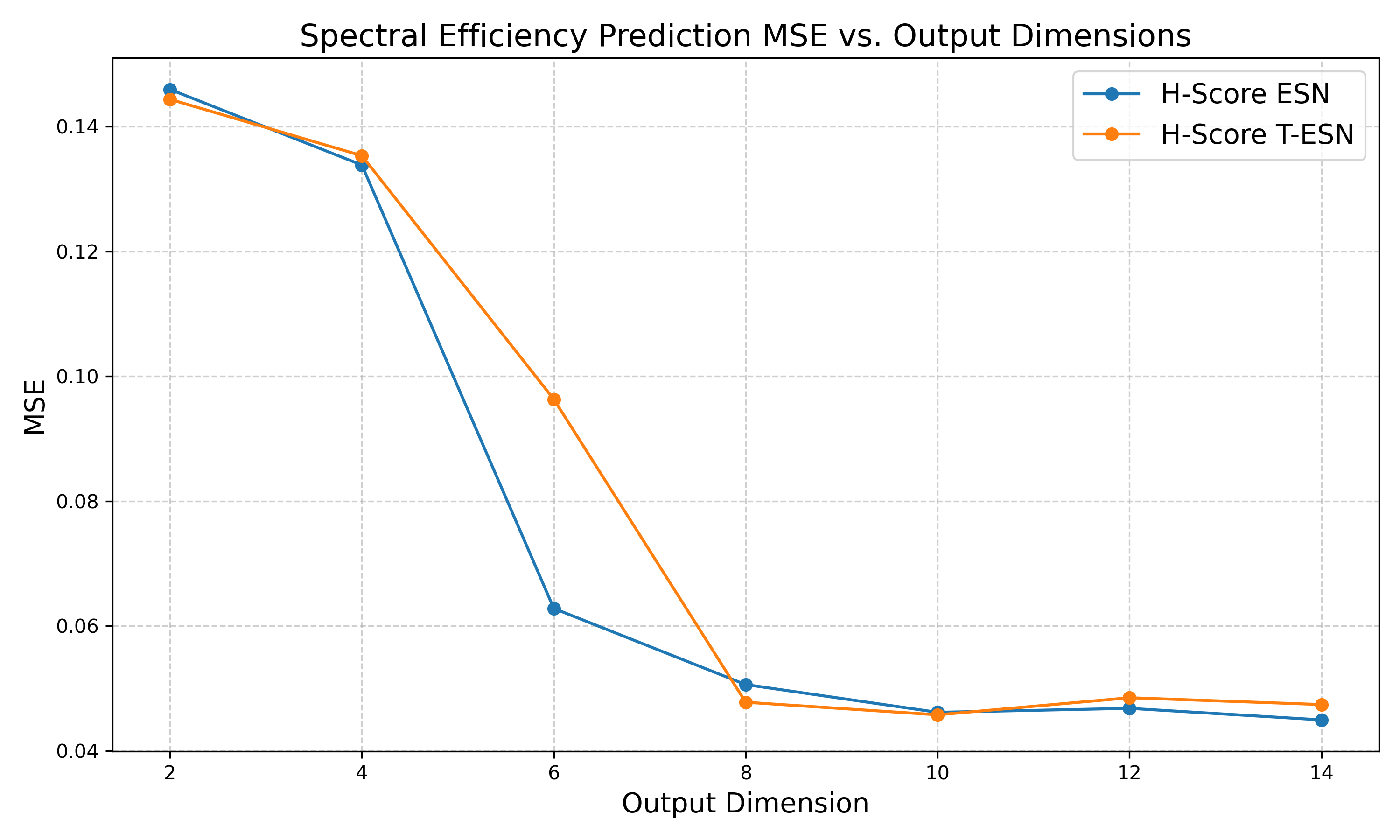} 
      \label{fig:semse_vs_dim}
     \end{minipage}
    }
\caption{KPI Prediction MSE against output dimensions.}
\label{fig:mse_dim}
\end{figure}

To identify this balance, we empirically evaluated the impact of the latent dimension on prediction performance using the H-score based encoders. As illustrated in Figures~\ref{fig:mse_vs_dim} and \ref{fig:semse_vs_dim}, which plots the prediction MSE against varying output dimensions $n$, we observed diminishing returns. While MSE decreases sharply for initial increases in $n$, the improvement slows down beyond $n=8$. 
This indicates that $n=8$  captures most of the crucial predictive information. Therefore, we selected this as our target dimension, achieving a substantial reduction from $(28\times 13)$ features to $8$ ($>45\times$ reduction).



\subsubsection{Full Training Regime}
\label{sec:fulldata}

Under full training conditions (80\% of data for training), the MLP with full KPIs consistently achieves the lowest prediction error, particularly for RSRQ. This outcome is expected, as a fully trained predictor tends to utilize all the informational content of the inputs. According to the data processing inequality, processed low-dimensional representations inherently cannot contain more information than the original input data.

\begin{figure}[t!]
    \centering
    \subfigure[RSRQ prediction MSE]{
     \begin{minipage}[t]{0.4\linewidth}
      \centering
      \includegraphics[width=\linewidth]{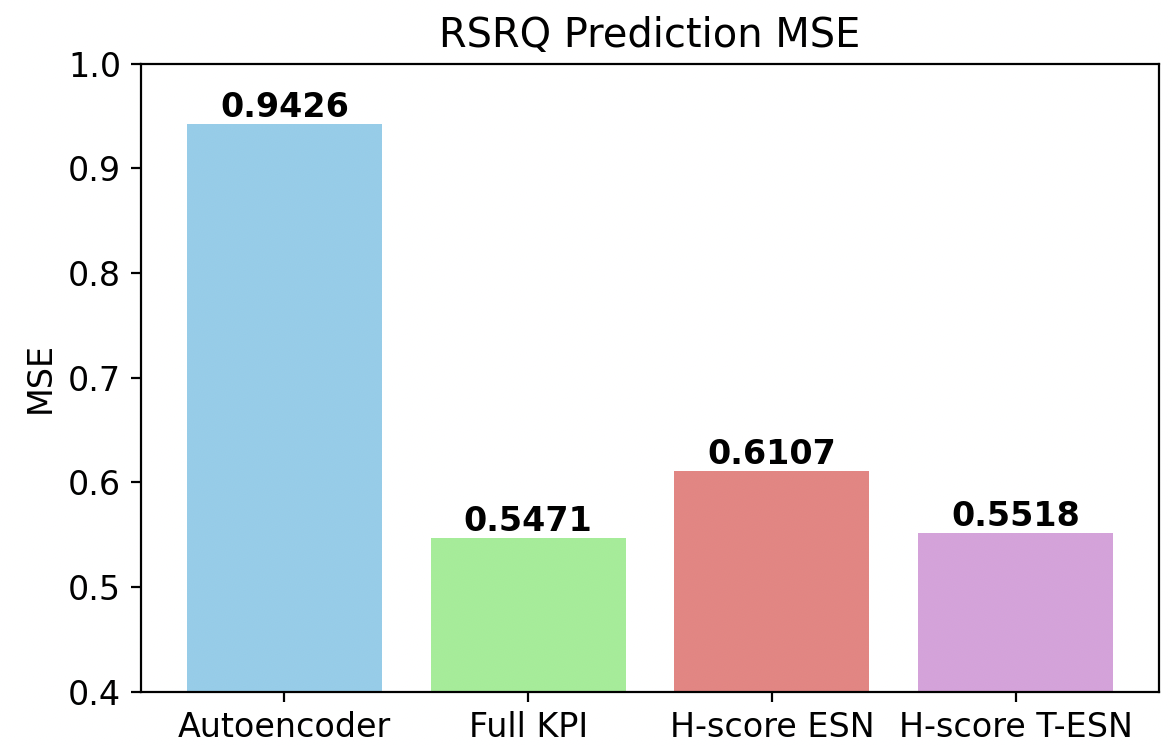} 
      \label{fig:rsrq_mse_20}
     \end{minipage}
    }
    \subfigure[RSRQ prediction correlation]{
     \begin{minipage}[t]{0.4\linewidth}
      \centering
      \includegraphics[width=\linewidth]{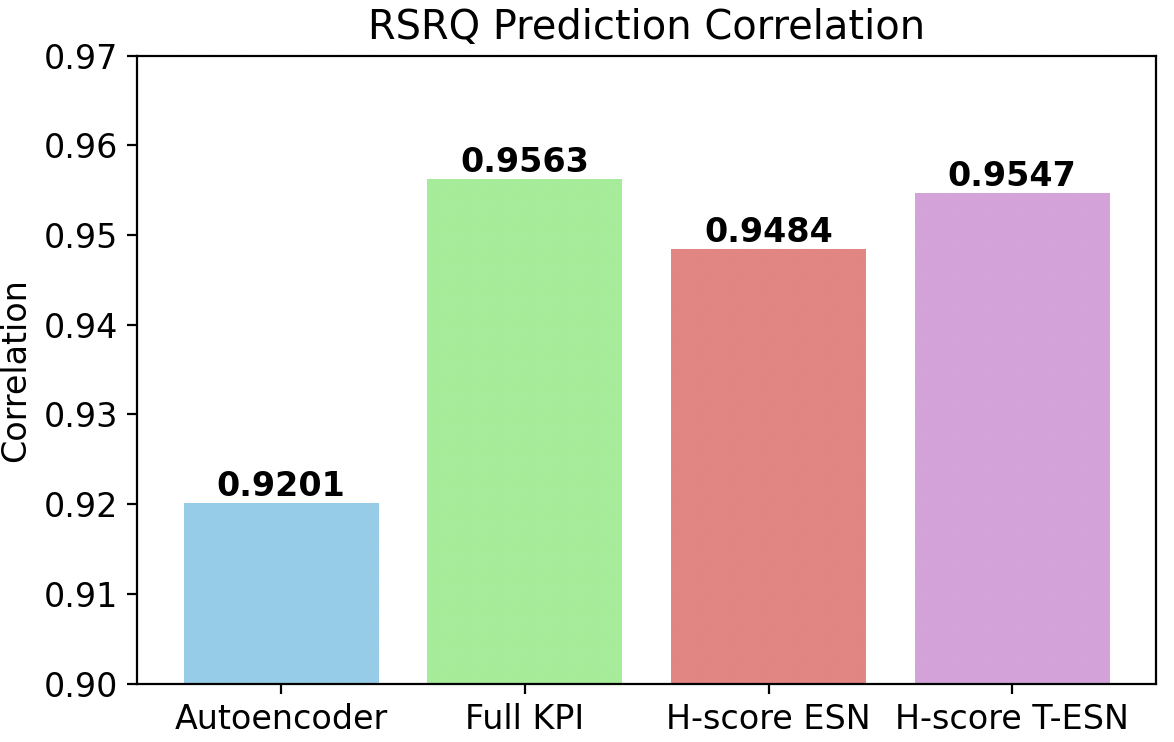} 
      \label{fig:rsrq_corr_20}
     \end{minipage}
    }
\caption{RSRQ prediction results under full training regime.}
\label{fig:rsrq_full}
\end{figure}

\begin{figure}[t]
    \centering
    \subfigure[Spectral efficiency prediction MSE]{
     \begin{minipage}[t]{0.4\linewidth}
      \centering
      \includegraphics[width=\linewidth]{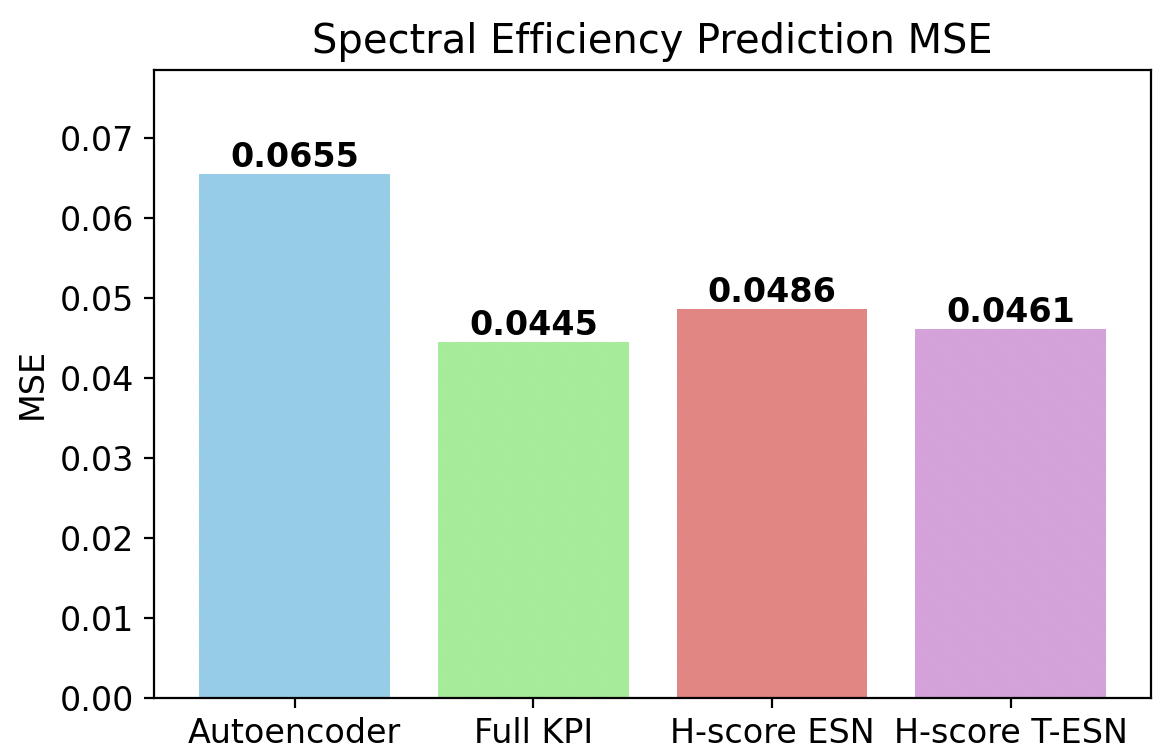} 
      \label{fig:se_mse_20}
     \end{minipage}
    }
    \subfigure[Spectral Efficiency prediction correlation]{
     \begin{minipage}[t]{0.4\linewidth}
      \centering
      \includegraphics[width=\linewidth]{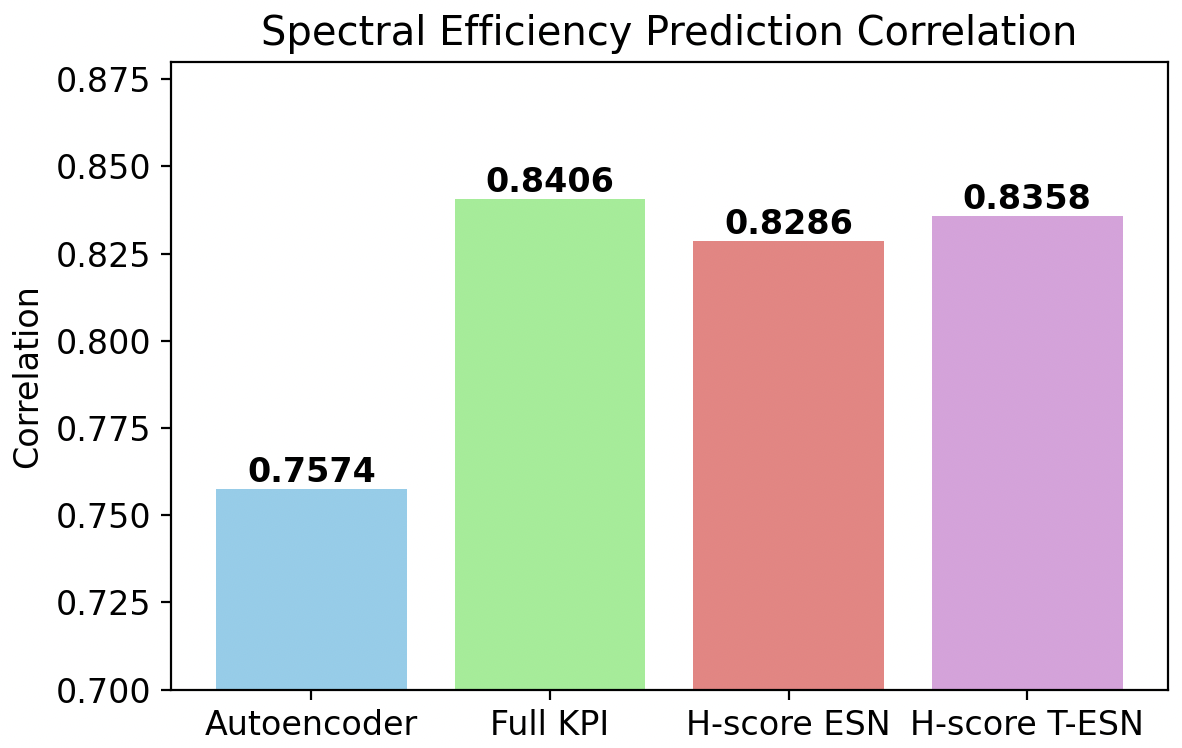} 
      \label{fig:se_corr_20}
     \end{minipage}
    }
\caption{Spectral efficiency prediction results under full training regime.}
\label{fig:se_full}
\end{figure}

As shown in Figures~\ref{fig:rsrq_full} and \ref{fig:se_full}, the green bars representing the MLP with full KPIs consistently achieve the lowest error, outperforming all other approaches across metrics. In contrast, the blue bars, representing the Autoencoder, exhibit the highest prediction error. The red bars and purple bars correspond to the H-score-trained ESN and T–ESN encoders respectively. 
Despite using fewer training epochs, both H-score based networks (red and purple) achieve results that are much better than the Autoencoder and competitive to the MLP with full KPIs  (green), demonstrating that their learned representations retain relevant information.
Quantitatively, the T–ESN achieves an MSE 0.8\% higher than the MLP with full KPIs for RSRQ prediction, and 3.6\% higher for spectral efficiency. Despite using significantly fewer epochs and a low-dimensional representation, the performance remains close to the MLP with full KPIs.


\subsubsection{Limited Data and Time Regime}

To simulate a constrained deployment scenario, we repeat the prediction task using only 5\% of the data for training and limit all models to 5 epochs. This 5\% subset corresponds to about 3000 training samples, while the remaining samples are reserved for testing. This environment reflects a realistic scenario with limited labeled data, requiring that the evaluation generalize broadly.

\begin{figure}[t!]
    \centering
    \subfigure[RSRQ prediction MSE]{
     \begin{minipage}[t]{0.4\linewidth}
      \centering
      \includegraphics[width=\linewidth]{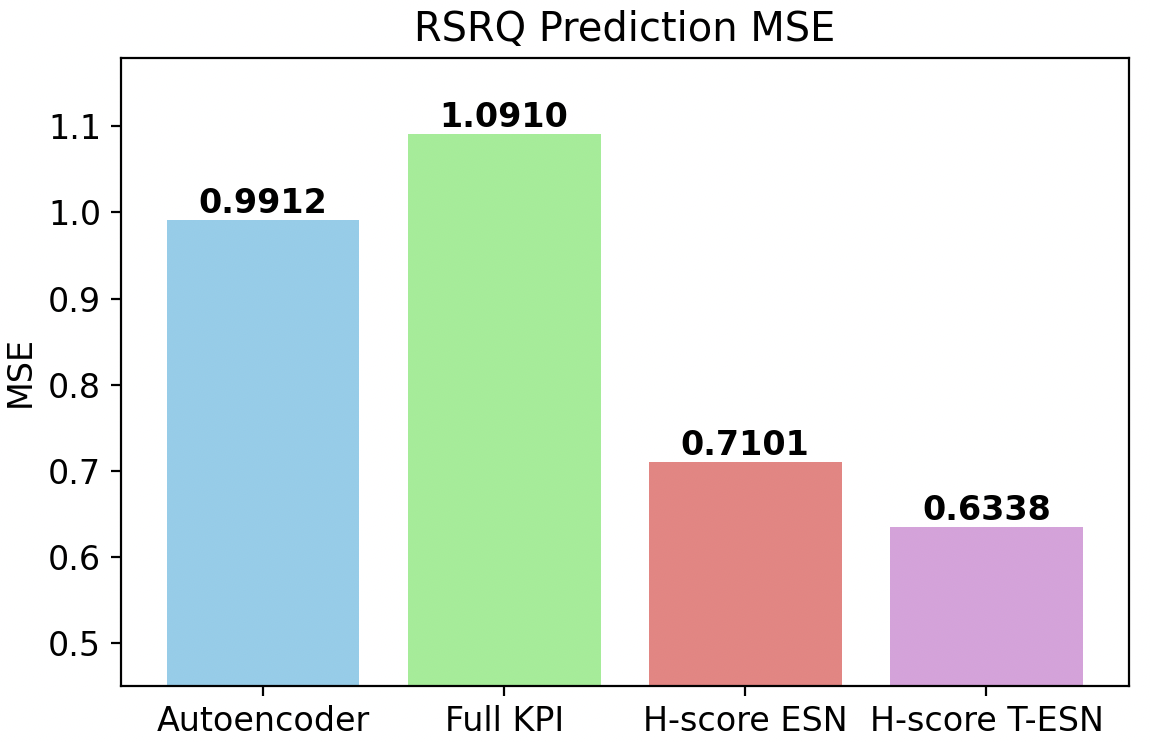} 
      \label{fig:rsrq_mse_95}
     \end{minipage}
    }
    \subfigure[RSRQ prediction correlation]{
     \begin{minipage}[t]{0.4\linewidth}
      \centering
      \includegraphics[width=\linewidth]{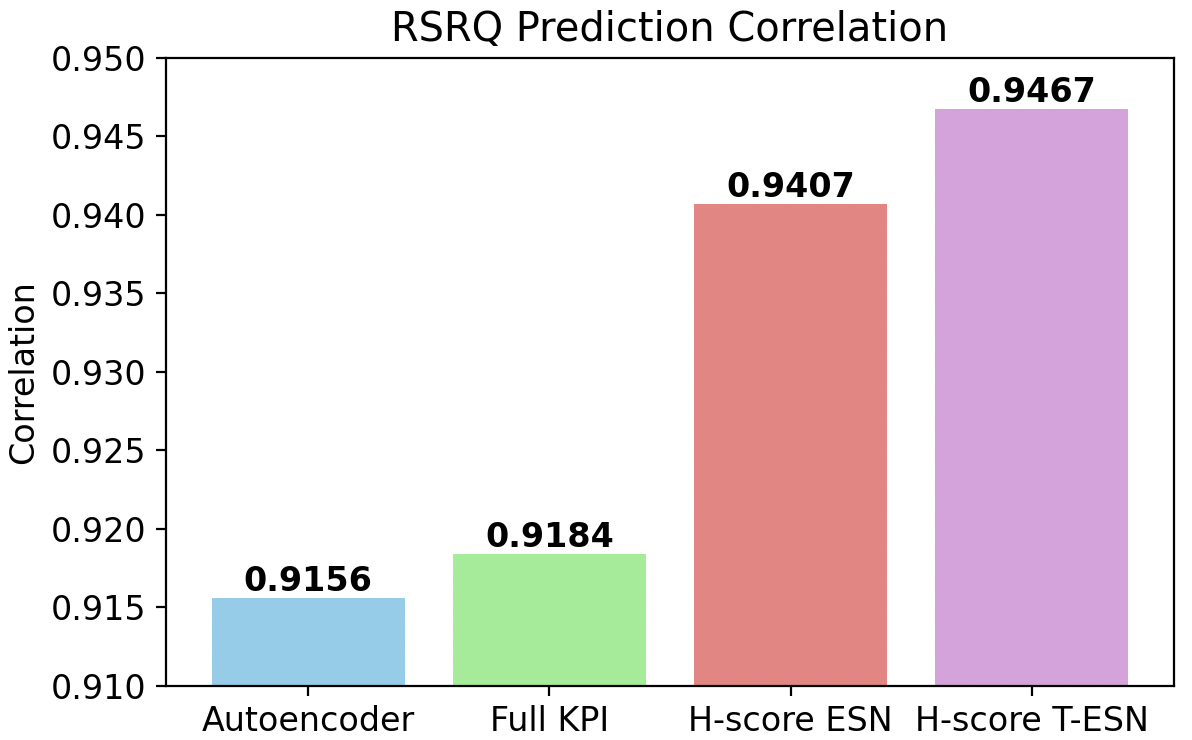} 
      \label{fig:rsrq_corr_95}
     \end{minipage}
    }
\caption{RSRQ prediction results with limited data (5\%) and training.}
\label{fig:rsrq-limited}
\end{figure}

\begin{figure}[t!]
    \centering
    \subfigure[Spectral efficiency prediction MSE ]{
     \begin{minipage}[t]{0.4\linewidth}
      \centering
      \includegraphics[width=\linewidth]{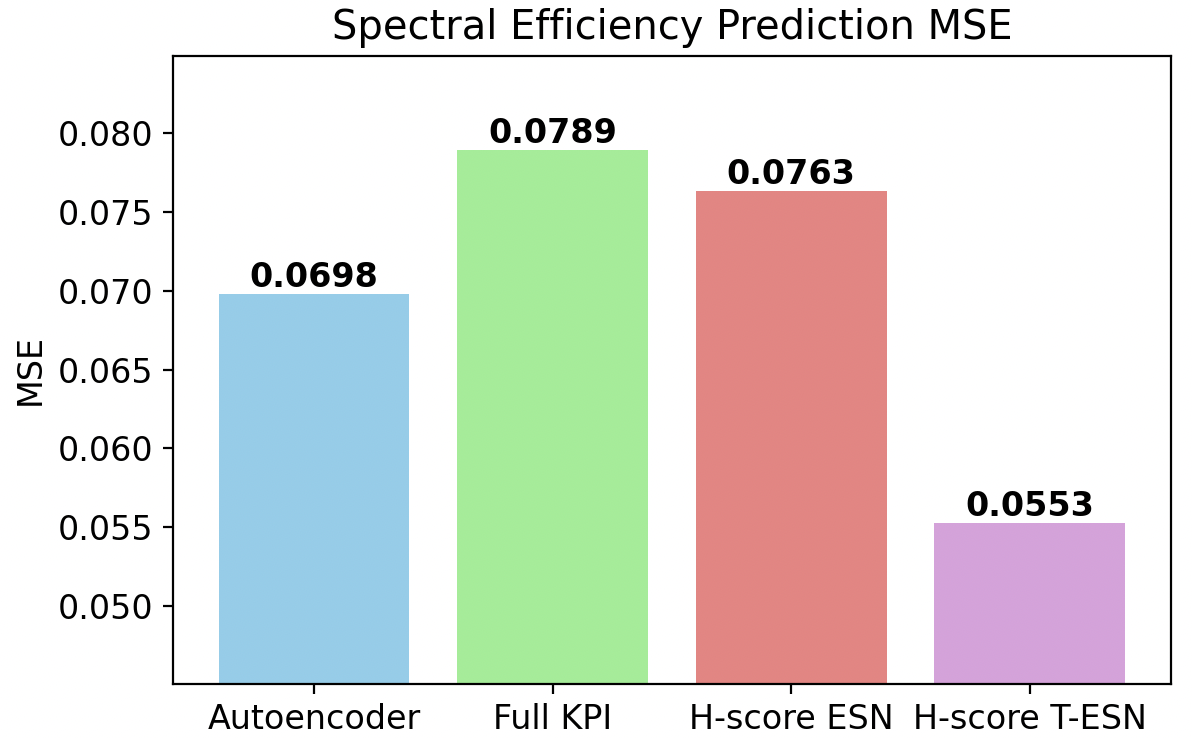} 
      \label{fig:se_mse_95}
     \end{minipage}
    }
    \subfigure[Spectral efficiency prediction correlation]{
     \begin{minipage}[t]{0.4\linewidth}
      \centering
      \includegraphics[width=\linewidth]{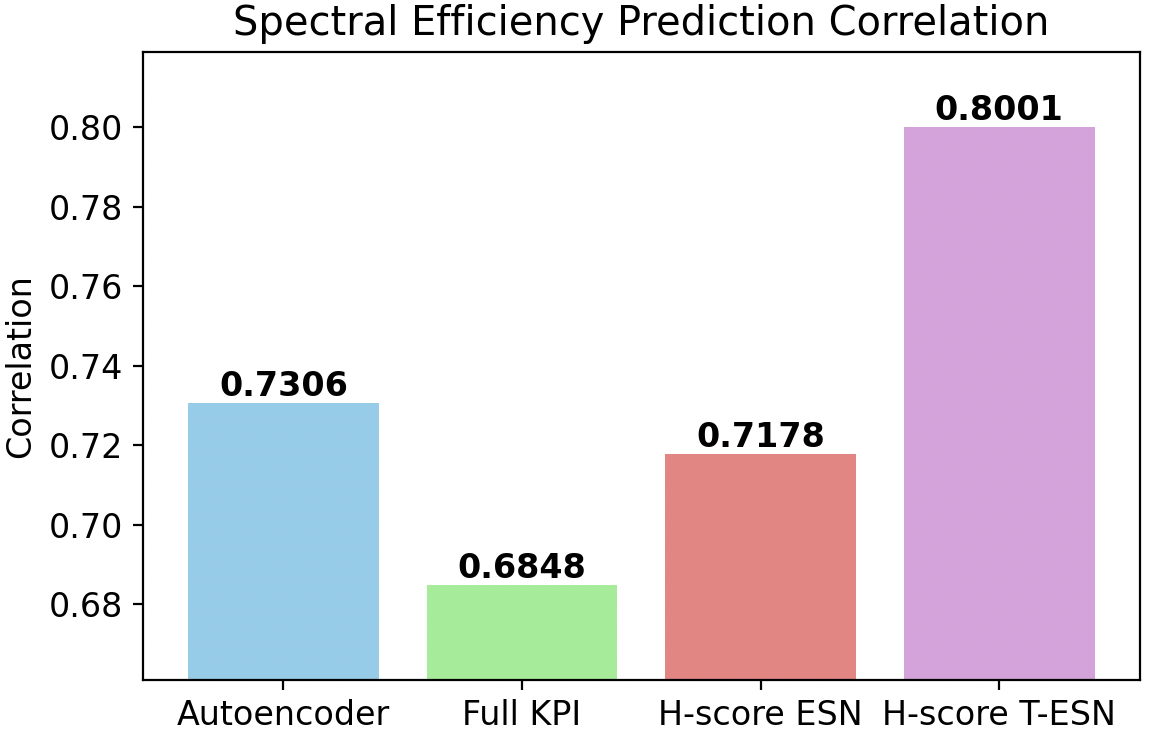} 
      \label{fig:se_corr_95}
     \end{minipage}
    }
\caption{Spectral efficiency results with limited data (5\%) and training.}
\label{fig:se-limited}
\end{figure}

As shown in Figures~\ref{fig:rsrq-limited} and \ref{fig:se-limited}, the performance trend reverses dramatically compared to the full training regime. Both models trained via H-score (red and purple) and Autoencoder now outperform the MLP with full KPIs (green) in spectral efficiency prediction.
In particular, the hybrid T-ESN (purple) emerges as the clear best performer, achieving the lowest error and highest correlation across both two target KPIs. Quantitatively, this translates to an advantage over the MLP with full KPIs trained under identical constraints: the T-ESN achieves a 41.9\% reduction in MSE for RSRQ and a 29.9\% reduction for spectral efficiency.

These results highlight the crucial advantage of task-aligned low-dimensional mapping when training resources are limited. 
This significant dimensionality reduction is key: it allows for a less complex predictor with fewer parameters. This simpler model translates to a lower sample complexity requirement, enabling the mapping models to learn efficiently and generalize better under harsher constraints. 
The T-ESN performs best, as its components working in tandem allow the learned representation to capture more meaningful information regarding complex system dynamics.

\section{Conclusions and Future Work}
\label{sec:concl}

We introduced a hybrid Transformer–ESN framework that learns compact, task-aligned representations of high-dimensional O-RAN KPIs via an H-score–driven objective. 
These embeddings serve as inputs to lightweight predictors that operate with reduced overhead and strong generalization.
Notably, the advantage of this architecture becomes most apparent in resource-constrained settings. By projecting high-dimensional KPIs into a compact space, the model reduces the sample complexity required for effective learning. This enables robust prediction with fewer training samples, making it well-suited for practical deployment in low-data regimes.
Looking ahead, we plan to enhance this framework by incorporating online, real-time learning for continuous adaptation of the H-score embeddings and predictors. 

\section*{Acknowledgment}
The project is funded by the Public Wireless Supply Chain Innovation Fund under Award No. 51-60-IF012. Any opinions, findings, and conclusions or recommendations expressed in this publication are those of the authors and do not necessarily reflect the views of the NTIA.

\bibliographystyle{IEEEtran}
\bibliography{IEEEabrv, ref}

\end{document}